\documentclass[aip,rsi,twocolumn, reprint]{revtex4-1} 
\usepackage{graphicx,amsmath,amssymb, url}
\usepackage{natbib}

\begin{document}
\bibliographystyle{abbrvnat}
\raggedbottom
\setlength{\parskip}{0pt}
\emergencystretch=1em
\Urlmuskip=0mu plus 1mu

\preprint{\\\texttt{\jobname.tex}}


\title{Fast Compact Laser Shutter Using a Direct Current Motor and 3D Printing} 

\author{Grace H. Zhang}
\email[]{ghzhang0@mit.edu}

\author{Boris Braverman}
\author{Akio Kawasaki}
\author{Vladan Vuleti\'{c} }

\affiliation{Department of Physics, MIT-Harvard Center for Ultracold Atoms and Research Laboratory of Electronics, Massachusetts Institute of Technology, Cambridge, Massachusetts 02139, USA}

\date{\today}

\begin{abstract}
We present a mechanical laser shutter design that utilizes a DC electric motor to rotate a blade which blocks and unblocks a light beam. The blade and the main body of the shutter are modeled with computer aided design (CAD) and are produced by 3D printing. Rubber flaps are used to limit the blade's range of motion, reducing vibrations and preventing undesirable blade oscillations. At its nominal operating voltage, the shutter achieves a switching speed of (1.22 $\pm$ 0.02) m/s with 1 ms activation delay and $10\;\mathrm{\mu s}$ jitter in its timing performance. The shutter design is simple, easy to replicate, and highly reliable, showing no failure or degradation in performance over more than $10^8$ cycles. 
\end{abstract}


\maketitle

Fast and robust optical shutters are essential to many laser-based experiments, as well as technological applications. These experiments, spanning a wide range of research areas in science and engineering, require the shutter to have the following performance characteristics: short switching time, small activation delay, high timing precision or low jitter, low vibration and heat dissipation, an aperture size capable of accommodating the relevant beam sizes, high extinction ratio, laser power handling sufficient for the blocked beam, and a long operation lifetime. 

To be useful for precise laser experiments, shutters require excellent timing performance, with short activation delays, fast switching times, and low timing jitters. A short activation delay allows for the generation of short light pulses, and a fast switching time allows for generating pulses with well-defined edges. A small timing jitter allows precise pulse timing control and synchronization with other parts of the experimental apparatus. These three qualities are necessary for the shutter to quickly and accurately realize a desired laser intensity profile over time. 

A shutter can be operated based on a large variety of mechanisms. Shutters based on electro-optical and acousto-optical modulators are very fast and hence have excellent timing performance, but require careful alignment. Most importantly, these types of shutters are unable to provide full extinction and transmission of the incident light, which can only be achieved by mechanical shutters. Many commercial laser safety interlock and diaphragm shutters exist \cite{Thorlabs_Shutters, RT_Tech_Shutters}, but they are typically not designed for applications in optics experiments; these shutters do not possess the desired timing performance and produce high noise and vibrations during operation. Commercial shutters with the requisite high performance \cite{SRS_Shutters, Uniblitz_Shutters} have costs that are impractical for typical laboratories, which can require tens of shutters. 

\begin{figure}[tb]
	\includegraphics[width=1\columnwidth]{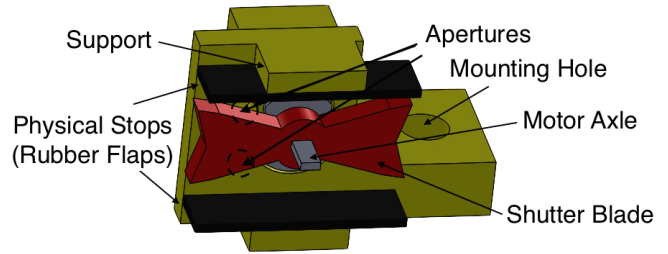} %
	\caption{CAD rendering of the shutter design. There is an upper aperture and a lower aperture, symmetric about the height of the motor axle, through which the laser beam can travel. The motor axle rotates the shutter blade to alternately block and unblock each aperture.} 
 	 \label{fig:solidworks}
\end{figure}

Lacking satisfactory commercial options, various laboratories have developed custom-made mechanical shutter systems \cite{Adams2000,Singer2002,Meyrath2003,Maguire2004,Mitchell2008,Martinez2011,Ye1990}. However, we find that these designs have drawbacks that limit their usability. Systems that utilize piezo actuators provide fast switching times and low timing jitter\cite{Adams2000}, but have insufficient extinction ratios and small sweep ranges. A design based on the modification of voice-coil motors in computer hard disk drives achieves both high speeds and a large aperture diameter \cite{Maguire2004}, but is too bulky in size for flexible placement in most optical setups. Another design, based on the thermal expansion of NiCr wire, achieves activation delay of less than 300 $\mathrm{\mu s}$ \cite{Ye1990}, but cannot operate at repetition rates with switching intervals shorter than 5 seconds due to the long relaxation time required for the thermal recovery of the wire. 

Most importantly, the aforementioned shutter designs capable of high speed and precision all require complicated and delicate assembly. This leads to three major disadvantages that severely restricts the large-scale application of these shutter designs. The difficult assembly process makes it hard to construct and maintain the shutters in the quantities typically required by a laboratory. Higher complexity decreases the reliability of the shutter as the overall shutter system contains more elements vulnerable to mechanical or electrical malfunctioning; this is especially important for applications that require long-term operation. Lastly, the high sensitivity of each shutter to fine variations in its construction process causes performance inconsistency among separate shutter units.

In this article we present a compact and high-performing shutter system that is simple to construct and easy to replicate consistently. The main components of the shutter are the 3D printed plastic mount and shutter blade, a DC electric motor, and an electrical driver circuit. Assembly of the shutter and its driver takes under an hour when provided the required materials, which consist of easily obtainable electrical components and 3D printed ABS plastic structures that can be produced in any desired quantity in one operation. Testing of the design shows both high speed and high reliability; the shutter has operated with sub-millisecond switching time for over $10^8 $ cycles without degradation. 

\begin{figure}[tb]
	\includegraphics[width=1.0\columnwidth]{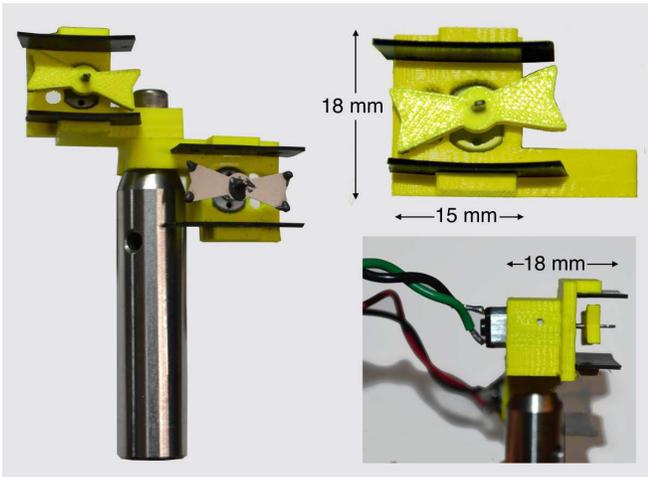} 
	\caption{Shutter units stacked on an optical post (left) and front and side view of shutter with plastic blade (right).} 
 	 \label{fig:photo}
\end{figure}

The shutter design is presented in Fig. ~\ref{fig:solidworks}. The DC motor rotates the shutter blade which blocks and unblocks the laser beam. The axle of the motor is flattened such that its rectangular cross section matches the shape of the hole in the blade through which the axle is inserted; this prevents slipping between the axle and the blade during operation. The axle is further secured in place by Loctite 420 cyanoacrylate adhesive (CA). Two neoprene rubber flaps act as physical stops to halt the motion of the blade. The rubber flaps are attached to the plastic support by CA and dampen vibrations from the blade that would otherwise cause the blade to bounce back and lead to undesirable transient blocking or unblocking of the beam. Laser beams can travel through the upper or lower aperture, both with a diameter of 2.5 mm, which are alternately blocked and unblocked during operation. The blade has a sweep range of 12 mm. The entire structure can be secured to an optical post through the mounting hole, as shown in Fig. ~\ref{fig:photo}, and is compact enough to be inserted into any optical setup.

\begin{figure}[tb]
	\includegraphics[width=0.9\columnwidth]{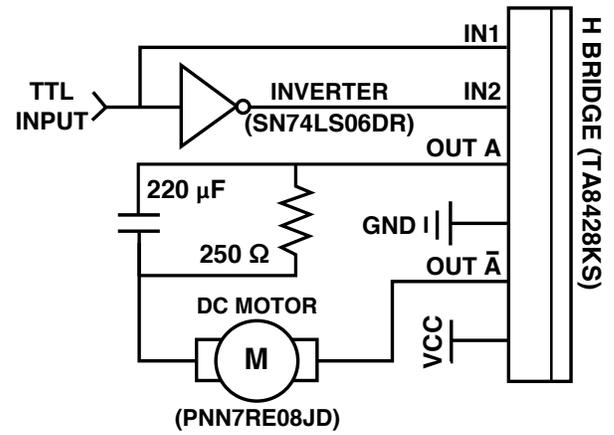}
	\caption{Circuit schematic for a single shutter, with part numbers indicated.} 
 	 \label{fig:schematic}
\end{figure}

The electrical circuit of a single shutter driver, presented in Fig. ~\ref{fig:schematic}, primarily consists of an H bridge and a pulse-generating RC circuit similar to another reported design \cite{Mitchell2008}. The circuit allows for bipolar driving of the motor with a unipolar electrical supply. The timing sequence of the opening and closing of the shutter is controlled by the TTL input signal.  

\begin{figure}[tb]
	\includegraphics[width=1\columnwidth]{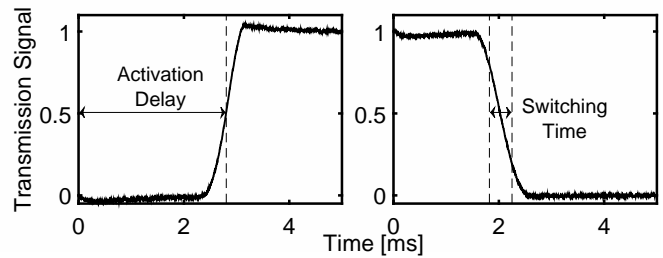}
	\caption{Transmission of laser beam (with 1 mm waist) through the aperture versus time elapsed after the TTL signal. Activation delay is the time interval between the application of the  TTL signal and 50$\%$ transmission. Switching time is the time interval between 80$\%$ transmission and 20$\%$ transmission.} 
 	 \label{fig:waveform}
\end{figure}

\begin{figure}[tb]
	\includegraphics[width=1\columnwidth]{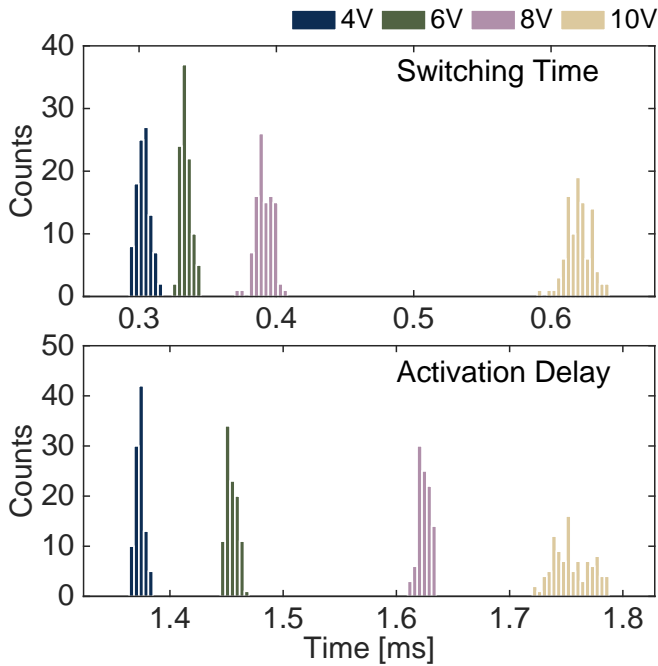}
	\caption{Distribution of 100 measurements for the switching times (top) and delay times (bottom) at different operating voltages $\mathrm{V_{CC}}$.} 
 	 \label{fig:VoltHist}
\end{figure}

\begin{figure}[tb]
	\includegraphics[width=1\columnwidth]{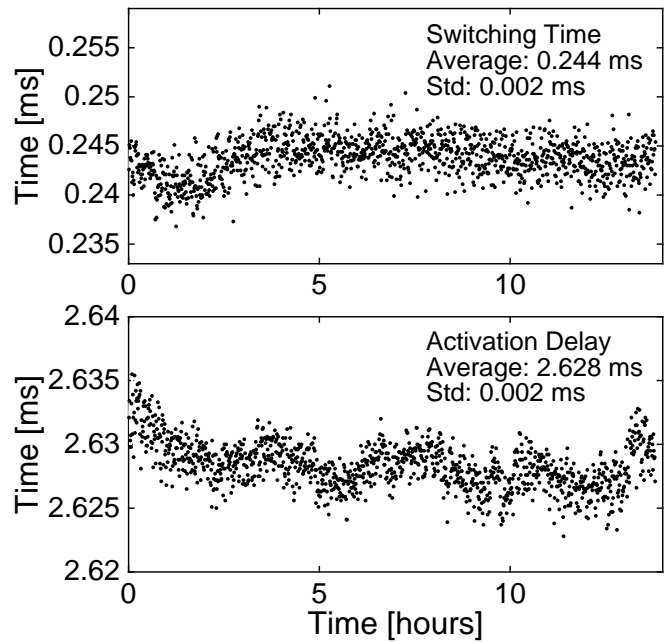}
	\caption{Jitter range and distribution of delay times (top) and switching times (bottom) over 14 hours with operating voltage $\mathrm{V_{CC} = 9\;V}$.} 
 	 \label{fig:jitter}
\end{figure}

This shutter design demonstrates excellent timing performance through its short activation delay, fast switching time, and low timing jitter. A fast avalanche photodiode (APD) was used to measure the transmission of a laser beam through the aperture during repeated blade transitions, giving the activation delay and the switching time of the shutter. Fig. ~\ref{fig:waveform} shows typical measurements of the transmission power of the laser beam during a single transition. Opening and closing delay times are between 1 and 2 ms, depending on the alignment of the laser beam through the shutter aperture. The speed of the shutter blade was measured by clipping an expanded laser beam with a metal sheet and determining the delay time taken by the shutter to provide complete extinction of the incident laser beam. This corresponds to the time between when the blade starts moving and when it reaches the edge of the metal sheet. By precisely translating the metal sheet parallel to the shutter blade's motion and recording the delay until complete extinction, the speed of the shutter blade was found to be (1.22 $\pm$ 0.02) m/s with the nominal operating voltage of 9 V, and increases as the operating voltage increases. As such, the activation delay and switching time are inversely correlated with the operating voltage, as shown in Fig. ~\ref{fig:VoltHist}. The typical jitter range in activation delay and switching time were measured to be about 20 $\mathrm{\mu s}$ and 10 $\mathrm{\mu s}$, respectively, also decreasing with increased operating voltage. The jitter of the shutter's timing response over 1000 measurements in a sample period of 14 hours are shown in Fig. ~\ref{fig:jitter}, illustrating the high mechanical stability provided by rubber flaps in arresting the motion of the shutter blade.

The shutter shows high consistency in the timing performance between each unit. The performance variation between different shutter units is smaller than the performance variation caused by removing the shutter from the setup and placing it back in the same position. 

The shutter presented in this design operates with low power usage, thereby extending its functional lifetime and minimizing its effects on the surrounding components in the optics setup through heating. The mechanical motion of the shutter is bistable; the shutter requires only 2-3 mA of current to keep the blade in its raised or lowered position, in contrast to unistable shutter systems which require current on the order of 100 mA to keep the shutter's position fixed \cite{Maguire2004, Singer2002}. The shutter has an operational voltage range of 4 to 15 V. Below 4 V, the motor cannot overcome static friction to actuate the blade motion, while above 15 V, excessive driving force causes the blade to bounce back from the rubber flap and cause temporary unwanted blockage or opening of the aperture. At an operating voltage of 9 V, the circuit only dissipates 13 mJ of energy in the shutter motor during each transition.

The shutter operates with a maximum short-term repetition rate of 110 Hz, and a maximum long-term repetition rate of 20 Hz. These limitations are due to the softening of the ABS plastic from high heat dissipation by the motor and could be solved by adding a heat sink to the mounting structure or by fabrication of the shutter components from more heat-resistant materials. The laser power-handling limit of the shutter with the plastic blade is 50 mW for a beam size of 1 mm. However, the plastic blade can be replaced with metal, as shown in Fig. ~\ref{fig:photo}, and the metal surface could be painted black to absorb and dissipate energy from the laser. We estimate that this would increase the power-handling limit to above 1 W. 

The shutters, both those with plastic blades and those with metal blades, produce minimal noise and vibrations. At an operating voltage of 4 V, operation is not detectable by ear. Noise and vibration could be further reduced by adding electronic braking to the driver circuit.

Applications of this mechanical shutter design can be further extended by replacing the opaque blade with an optical component. In this way, one can remotely place components such as a polarizing element, a wave plate, or an attenuator, in and out of the path of a laser beam. The low jitter in shutter timing is equivalent to a high repeatability of the blade position during each shutter cycle. This suggests that even alignment-sensitive elements such as lenses or prisms could also be used in place of the opaque blade for controlling laser beam alignment or focusing.

In conclusion, we have designed and implemented a simple, compact, robust, low-noise, high-performance laser shutter, based on a small commercial DC motor, 3D printed shutter body and blade, and rubber motion stops. The shutter has a lifetime and timing performance comparable to commercial shutters with both very low cost and time of construction. Most importantly, this shutter is easy to reproduce in large numbers with high consistency, as the mechanical structure is fixed by the CAD file used for 3D printing and requires no extra tinkering during assembly, and is highly reliable due to the elimination of variations in assembly and the simplicity of the mechanism. We expect this design to be highly useful in laser experiments across many disciplines, where it could fill a niche where commercial options are currently limited. 

\begin{acknowledgments}
This research was supported by NSF, DARPA QuASAR, and by MURIs through AFOSR and ARO. G.Z. acknowledges support from the Undergraduate Research Opportunity Program at MIT and B.B. from the National Science and Engineering Research Council of Canada.
\end{acknowledgments}

\bibliography{ShutterPaper}

\begin{thebibliography}{11}
\providecommand{\natexlab}[1]{#1}
\providecommand{\url}[1]{\texttt{#1}}
\expandafter\ifx\csname urlstyle\endcsname\relax
  \providecommand{\doi}[1]{doi: #1}\else
  \providecommand{\doi}{doi: \begingroup \urlstyle{rm}\Url}\fi

\bibitem[Adams(2000)]{Adams2000}
C.~Adams.
\newblock A mechanical shutter for light using piezoelectric actuators.
\newblock \emph{Review of Scientific Instruments}, 71\penalty0 (1):\penalty0
  59--60, 2000.

\bibitem[Maguire et~al.(2004)Maguire, Szilagyi, and Scholten]{Maguire2004}
L.~Maguire, S.~Szilagyi, and R.~Scholten.
\newblock High performance laser shutter using a hard disk drive voice-coil
  actuator.
\newblock \emph{Review of Scientific Instruments}, 75\penalty0 (9):\penalty0
  3077--3079, Sep 2004.
\newblock ISSN 0034-6748.
\newblock \doi{10.1063/1.1786331}.

\bibitem[Martinez et~al.(2011)Martinez, Hernandez, Reyes, Gomez, Ivory,
  Davison, and Aubin]{Martinez2011}
S.~Martinez, L.~Hernandez, D.~Reyes, E.~Gomez, M.~Ivory, C.~Davison, and
  S.~Aubin.
\newblock Note: Fast, small, and low vibration mechanical laser shutters.
\newblock \emph{Review of Scientific Instruments}, 82\penalty0 (4):\penalty0
  046102, 2011.
\newblock \doi{http://dx.doi.org/10.1063/1.3574224}.
\newblock URL
  \url{\\\url{http://scitation.aip.org/content/aip/journal/rsi/82/4/}\\\url{10.1063/1.3574224}}.

\bibitem[Meyrath(2003)]{Meyrath2003}
T.~P. Meyrath.
\newblock Inexpensive mechanical shutter and driver for optics experiments.
\newblock \url{http://george.ph.utexas.edu/~meyrath/informal/shutter.pdf},
  September 2003.
\newblock URL \url{http://george.ph.utexas.edu/~meyrath/informal/shutter.pdf}.

\bibitem[Mitchell and Lebel(2008)]{Mitchell2008}
D.~Mitchell and P.~Lebel.
\newblock Ultrafast mechanical shutters for laser cooling applications: The
  ishutter system.
\newblock
  \url{http://www.phas.ubc.ca/~qdg/publications/InternalReports/LM-APSC479.pdf},
  January 2008.
\newblock URL
  \url{http://www.phas.ubc.ca/~qdg/publications/InternalReports/LM-APSC479.pdf}.

\bibitem[{RT Technologies}()]{RT_Tech_Shutters}
{RT Technologies}.
\newblock Laser beam shutters.
\newblock \url{http://www.rtlasersafety.com/laser-beam-shutters.php}.

\bibitem[Singer et~al.(2002)Singer, Jochim, Mudrich, Mosk, and
  Weidem{\"u}ller]{Singer2002}
K.~Singer, S.~Jochim, M.~Mudrich, A.~Mosk, and M.~Weidem{\"u}ller.
\newblock Low-cost mechanical shutter for light beams.
\newblock \emph{Review of Scientific Instruments}, 73\penalty0 (12):\penalty0
  4402--4404, 2002.
\newblock \doi{http://dx.doi.org/10.1063/1.1520728}.
\newblock URL
  \url{http://scitation.aip.org/content/aip/journal/rsi/73/12/10.1063/1.1520728}.

\bibitem[{Stanford Research Systems}()]{SRS_Shutters}
{Stanford Research Systems}.
\newblock Laser shutter systems {SR470 \& SR474} - laser shutters \&
  controllers.
\newblock \url{http://www.thinksrs.com/products/SR470474.htm}.
\newblock URL \url{http://www.thinksrs.com/products/SR470474.htm}.

\bibitem[{Thorlabs}()]{Thorlabs_Shutters}
{Thorlabs}.
\newblock Optical shutter.
\newblock \url{https://www.thorlabs.us/newgrouppage9.cfm?objectgroup_id=927}.

\bibitem[{Vincent Associates}()]{Uniblitz_Shutters}
{Vincent Associates}.
\newblock {LS2} 2mm uni-stable shutters.
\newblock \url{https://www.uniblitz.com/product/ls2-shutter-system/}.
\newblock URL \url{https://www.uniblitz.com/product/ls2-shutter-system/}.

\bibitem[Ye et~al.(1990)Ye, Jiang, and Wong]{Ye1990}
M.~Ye, D.~Jiang, and C.~Wong.
\newblock Simple electromechanically operated optical shutter.
\newblock \emph{Review of Scientific Instruments}, 61\penalty0 (7):\penalty0
  2003, 1990.

\end{thebibliography}

\end{document}